\documentclass[twocolumn,twoside,slac]{revtex4}
\usepackage{graphicx}
\usepackage{fancyhdr}
\begin{document}

\title{Grid Based Monitoring on the Rutgers CDF Analysis Farm}

%

\author{P. Jacques, F. Ratnikov, T. Watts}
\affiliation{Rutgers, the State University of New Jersey, Piscataway, NJ 
08854-8019, USA}
\author{I. Terekhov}
\affiliation{FNAL, Batavia, IL 60510, USA}

\begin{abstract}
Run II at the Fermilab Tevatron Collider started in March 2001, and it
 will continue probing the high energy frontier in particle physics until
 the start of the LHC at CERN. The CDF collaboration at Fermilab has already
 stored 260 TB of data and expects to store 1PB of data in the next
 two years. The HEXCAF computing farm is being set up at Rutgers University to
 provide the software environment, computing resources, and access to data
 for physicists participating in the Collaboration. Some job submission,
 detector data access and storage of the output results are based on the
 SAM-GRID tools. To extend monitoring for these jobs running on the farm a 
bridge was
 developed between the SAM-GRID monitoring tools and the internal farm 
monitoring.
 This presentation will describe the configuration and functionality of the 
HEXCAF farm
 with the emphasis on the monitoring tools.
 Finally we summarize our experience of installing and operating a GRID 
environment on a remote
 cluster that is being used for real physics studies in the big running 
experiment.
\end{abstract}

\maketitle

\thispagestyle{fancy}


\section{HEX Farm at Rutgers}
The HEX farm \cite{hexfarm} was developed for the {\bf H}igh Energy Physics 
{\bf EX}perimental
group of Rutgers University. The group contains about 20 physicists 
participating in CDF, CMS and other experiments, with
about 70
The primary goal of the farm is to satisfy computing needs of the CDF group 
for physics data analysis.

HEX farm is a good example of a small analysis farm, as opposed to a big 
production farm,
which serves a typical university group participating in a big HEP experiment.

\section {Hardware configuration}

The farm is based on dual CPU PCs running Fermi Linux \cite{frh}. Due to lack 
of IP addresses on the university campus,
15 worker nodes\footnote {by conference time, recently upgraded to 23 workers} 
are connected in a private
network. Big IDE disks connected to the worker nodes are cross mounted on 
every node of the farm
\footnote {by conference time, a recently dedicated file server was also 
installed}.

Three special ''portal'' nodes are connected to both WAN and private networks:
\begin {itemize}
\item Farm header node. Users here interactively prepare and test their jobs. 
It also hosts home directories
  and experiment specific software, and runs NAT and NIS servers.
\item Data transfer node. This is used to pass data between public and private 
networks. Traffic is mainly data
  exchange between Rutgers and Fermilab; this channel network bandwidth is 
about 100 Mbits/s.
\item Analysis facility header node. It controls analysis load, runs 
SAM\footnote {more details about SAM in \ref{sam}}, and
  hosts SAM data cache. This is the only fully kerberized node on the farm. It 
is connected to the Fermilab kerberos domain.
\end {itemize}

As most of the jobs running on the farm are related to the CDF experiment, the 
software and computer setup
 is mostly inherited from standard CDF installations\cite{cdf_software}.

\section {Resource Management}

To minimize compatibility issues, the HEX farm uses CDF approved components
for resource management. 
CPU usage and data storage are farm resources essential to manage.

 We installed batch system developed for and used on the CDF Central Analysis 
Facility (CAF)
\footnote {more details about CAF in \ref{caf}}\cite{caf_conf}. 
The system is built on top of the FBSNG batch system \cite{fbsng} developed at 
Fermilab.
 
The HEX farm does not have its own mass storage system. Necessary data either 
sits in the local disk cache, some data locally created and some
transferred from Fermilab,  or can be  accessed  from the CDF central data 
handling system at Fermilab directly over the WAN.

Since the CDF Data Handling system is evolving\cite {dcache}, there are 
different data delivery and cataloging systems
in use, both at the Fermilab and at Rutgers.

\subsection {Disk Inventory Manager (DIM) and Data File Catalog (DFC)}
DIM is the original light weight CDF data cache system\cite {dim}. It is 
natively compatible with CDF software.
Central DIM running at Fermilab accesses the central mass storage.
The central DFC uses an Oracle database to store metadata; the HEX farm
instance uses MSQL. The
HEX farm instance of DIM is not connected to mass storage. Instead it can 
fetch data from
the central data handling system. A very convenient feature of the DIM at 
Rutgers is the ability to
synchronize a dataset replica with its master copy at Fermilab. It is 
important because CDF is a running
experiment and new data are continuously added to datasets. Datasets 
interesting to Rutgers physicists
are set up in such a way that when new files are added to the dataset at 
Fermilab and a user asks to read the dataset, the new files are automatically 
fetched
to Rutgers and local metadata are updated accordingly. dCache\cite{dcache} 
data access is used for data delivery. The data transfer unit in this mode is 
one fileset, that is about 10 one-GB files.

\subsection {Sequential Data Access via Metadata (SAM)\label{sam}}
SAM \cite{sam1, sam2, sam3} is a sophisticated baseline data handling solution 
for the D0 experiment at Fermilab.
CDF is currently adopting SAM. This system uses a local data cache and 
centrally located metadata. Required
data are delivered automatically from the closest available source including 
the central mass storage at Fermilab.
SAM provides a native mechanism for cataloging and storing in the central mass 
storage system of new data produced
on the local site. 

SAM is still in trial use by CDF and is installed on several sites including 
the HEX farm. The SAM data cache
disks are located on the SAM station and exported to worker nodes via NFS.

\subsection {Job Submission \label{caf}}
The CAF batch system is used on the HEX farm.
User identity is authenticated via Kerberos through the collection of 
collaboration CAF systems. This is the reason that the HEX farm head node is
attached to the Fermilab kerberos domain.
The CAF system decouples well data submission, job executing, and output 
delivery:
the job  is remotely submitted to the farm, executed there, and output is 
delivered to the specified remote destination.

This  approach  provides a good possibility for distributed computing. From 
user point of view, all CAF farms are
identical providing user is authorized to submit jobs there. 
The same GUI can submit the same job to any of the available CAF farms. The 
following is assumed:
\begin{itemize}
\item baseline software is available on the farm
\item job tarball created on the fly contains all files needed to run in the 
base environment
\end{itemize}     
This system is naturally extendible for resource broker driven operation.

\subsection {GRID Approach}
The Hex farm has also installed a prototype of the Job and Information Manager 
(JIM) \cite{jim}.
This tool provides resource brokering, job submission, and a monitoring suite. 
The system is built
on top of Condor-G and uses SAM information about the availability of data 
files to select a destination for the job.
JIM has an adapter to the local CAF submitter and uses LDAP based information 
providers and monitoring tools.
The prototype operation was demonstrated at the SC2002 workshop, the HEX farm 
being one execution site.

\section {Monitoring}
Both JIM and CAF have nice monitoring systems. Each one provides
complete information about details of the system status, history
etc. But the connection between these monitoring domains was broken
when the JIM-CAF adapter submitted a job to the batch system. JIM
operates in terms of a global universal job ID and CAF identifies the
job by a local ID that is assigned when the job is released to the
batch queue. This problem is general for any GRID system submitting
jobs to local batch systems that are not ``gridifyed''. A database
driven solution was selected for JIM and is implemented on the HEX
farm.

\begin{figure}[h]
\centering
\includegraphics[width=65mm]{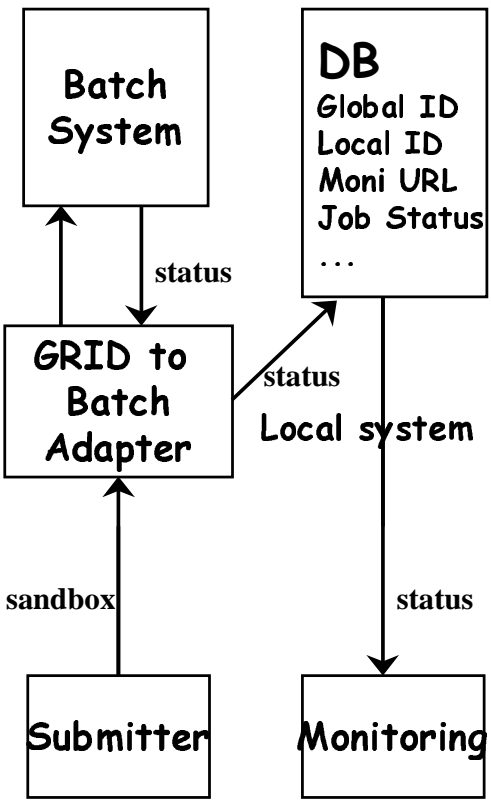}
\caption{Database driven approach for global monitoring of the local system} 
\label{monigraph}
\end{figure}

\begin{figure*}[h]
\centering
\includegraphics[width=150mm]{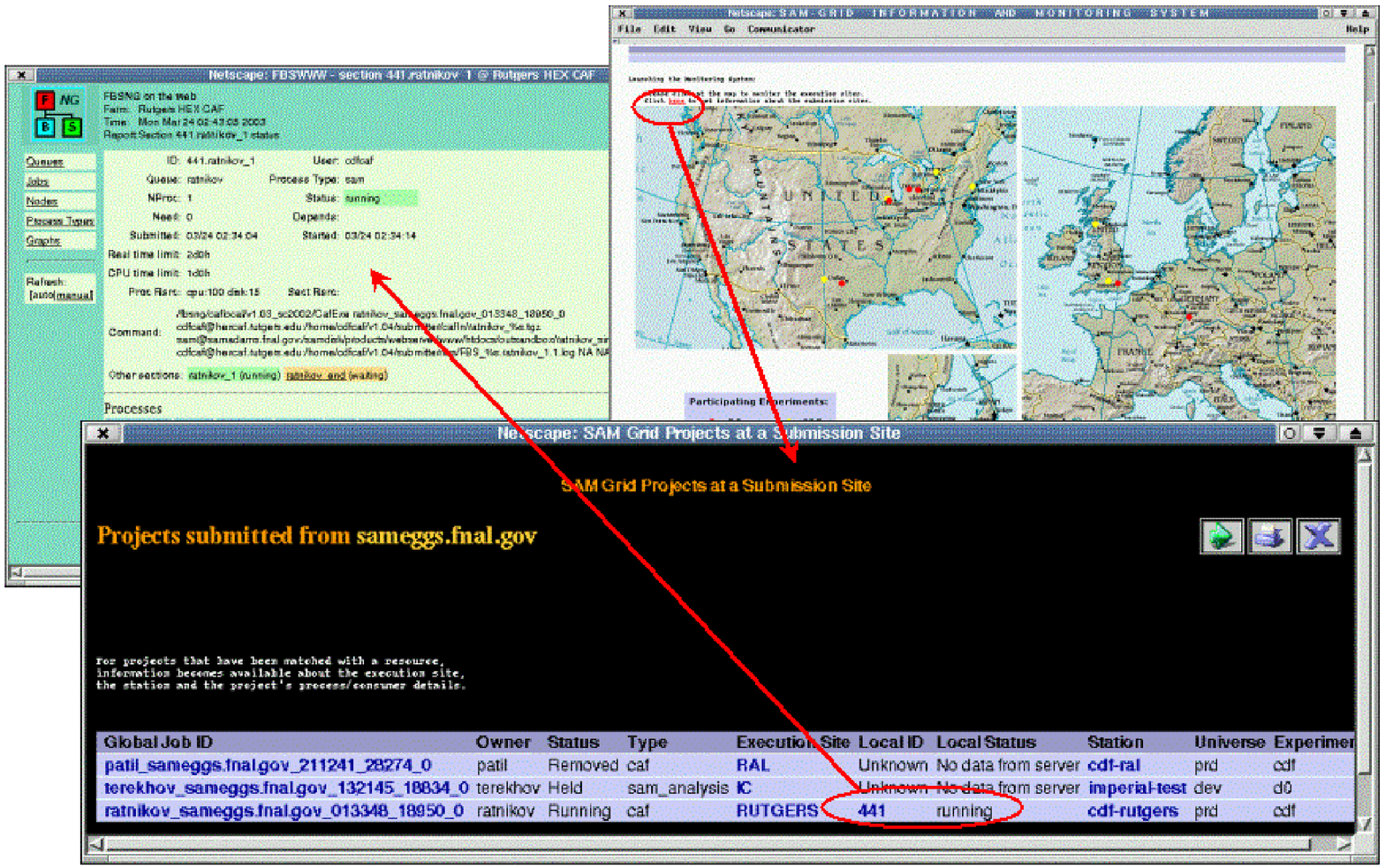}
\caption{Bridge between global and local monitoring suits. The list of all JIM 
jobs was originally available
(top right). The farm information server provides information about the local 
ID and job status for a global job
(bottom). This information allows the monitoring of a particular job on the 
CAF local monitoring page
(top left) and that gets further local details about the job 
operation.} \label{moni}
\end{figure*}

Figure \ref{monigraph} presents the method of monitoring. The job submitter 
delivers the sandbox to the GRID-Batch 
adapter on the local system. The adapter submits the job to the local queue 
and gets back the local ID of the job.
The pair of global job ID and local job ID is put into the database together 
with ongoing status
of the job. This database record works as a bridge between the global and 
local system and makes possible the
requesting of the status of the local job by global ID. After the job is 
completed, the corresponding record
may be removed from the database. A simple file based database was used to 
prove the principle and improve
monitoring tools\footnote {by conference time, an XML based database was
in use}.
Figure \ref{moni} demonstrates how information about matching between global 
and local IDs allows a user to follow the
local status of a global job. 
 
\section {Operation Experience}
In summary, the HEX farm has  a variety of different data access and job 
submission systems installed.
The farm is involved in essential CDF tests of frontier computing technologies.
 At the same time
it is operating as a computing facility for a university group. This is 
an interesting test that 
can demonstrate the choice by physicists who are not enthusiastic about new 
computer technologies
and who would rather use these computers to proceed with physics analysis. 
Experience shows that:
\begin{itemize}
\item 80\% of data are about 25 static DIM datasets resident on the disk (some 
replicas of Fermilab datasets and some locally created).
These data are intensively used for data analysis.
\item Nearly 100\% of big jobs are submitted via the standard CAF submitter.
\end{itemize}   

The farm is used for analysis, therefore neither massive data production nor 
huge Monte Carlo jobs
run there. Static datasets can easily be managed manually in the small 
community. Although SAM data
handling is very sophisticated, the DIM has been in use by CDF for a long time 
and is more familiar to people.
Our experience is that users prefer simple and reliable solutions. At the same 
time,
simple solutions are usually less scalable and more sophisticated ones become 
simpler for use
when the system grows.

The HEX farm already has one and half times more worker nodes and twice
more computing power, than at conference time, so massive Monte Carlo 
generation can be reasonably performed on site.
New generated data need to be transferred to  the central CDF mass storage. 
Then the automated SAM solution becomes
faster and simpler with respect to data transfer and cataloging on the central 
system; the
SAM approach will be definitely appreciated by HEX farm users.

Concluding, in running an analysis farm for the Rutgers HEP experimental group 
we contribute to the development and testing of 
innovative computing technologies deployed for the CDF collaboration and keep 
innovations tuned for use
by the physicists of the group.



\end{document}